\documentstyle[aaspptwo]{article}
\tighten          

\newcommand{\beq}{\begin{equation}}
\newcommand{\eeq}{\end{equation}}
\newcommand{\beqa}{\begin{eqnarray}}
\newcommand{\eeqa}{\end{eqnarray}}

\newcommand{\dbv}{$\Delta ({\rm B} - {\rm V})$}
\newcommand{\dv}{\hbox {$\Delta \rm V^{TO}_{HB}$}}
\newcommand{\dvtwo}{\Delta {\rm V}}

\newcommand{\mvrr}{\hbox {$\rm M_v(RR)$}}
\newcommand{\mvto}{\hbox {$\rm M_v(TO)$}}

\newcommand{\rgc}{\hbox {R$_{\rm GC}$}}
\newcommand{\ea}{{\it et al. }}
\newcommand{\feh}{\hbox{$ [{\rm Fe}/{\rm H}]$}}

\newcommand{\dh}{Debye-H\"{u}ckel}

\lefthead{Chaboyer et al.}
\righthead{Formation of the Galactic Halo}

\begin{document}

\title{Globular Cluster Ages and the Formation of the Galactic Halo}

\author{Brian Chaboyer}
\affil{Canadian Institute for Theoretical Astrophysics,
60 St.~George Street, Toronto, Ontario, Canada M5S 1A7~~E-Mail:
chaboyer@cita.utoronto.ca }

\author{P.~Demarque}
\affil{Department of Astronomy, and Center for Solar and Space
Research, Yale University, Box 208101, New Haven, CT 06520-8101
{}~~E-Mail: demarque@astro.yale.edu}

\and
\author{Ata Sarajedini}
\affil{Kitt Peak National Observatory, National  Optical Astronomy
Observatories\altaffilmark{1}, PO Box 26732, Tucson, AZ 85726~~E-Mail:
ata@noao.edu}

\altaffiltext{1}{NOAO is operated
by the Association of Universities for Research in Astronomy, Inc., under
contract with the National Science Foundation.}

\begin{abstract}

Main sequence turnoff magnitudes from the recent set of Yale
isochrones (Chaboyer \ea 1995) have been combined with a variety of
relations for the absolute magnitude of RR Lyr stars (\mvrr) to
calibrate age as a function of the difference in magnitude between the
main sequence turn-off and the horizontal branch (\dv).  A best estimate
for the calibration of \mvrr ~is derived from a survey of the current
literature: $\mvrr = 0.20\,\feh + 0.98$.  This estimate, together with
other calibrations (with slopes ranging from 0.15 to 0.30) has been
used to derive \dv ~ages for 43 Galactic globular clusters.
Independent of the choice of \mvrr, there is no strong evidence for an
age-Galacto\-centric distance relationship among the 43 globular
clusters.  However, an age-metallicity relation exists, with the
metal-poor clusters being the oldest.  A study of the age distribution
reveals that an age range of 5 Gyr exists among the bulk of the
globular clusters.  In addition, about 10\% of the sample are
substantially younger, and including them in the analysis increases
the age range to 9 Gyr.  Once again, these statements are independent
of the \mvrr ~relation.  Evidence for age being the second parameter
governing horizontal branch morphology is found by comparing the
average \dv ~age of the second parameter clusters to the normal
clusters.  The second parameter clusters are found to be on average 2
-- 3 Gyr younger than the other clusters, which is consistent with age
being the second parameter. These results suggest that globular
clusters were formed over an extended period of time, with
progressively more metal-rich globular clusters ($\feh \ga -1.7$)
being formed at later times.

\end{abstract}

\keywords{globular clusters: general -- Galaxy: formation -- Galaxy: Halo}

\section{Introduction}

Understanding the process of galaxy formation continues to be one of
the key quests in astrophysics.  In this regard, the Milky Way plays
a unique role because it is the only galaxy for which we can obtain
detailed chemical, kinematic and chronological information. Observational
and theoretical studies over the last 60 years have lead to a basic
understanding of how the Galaxy formed (see Larson 1991).  It is clear
that the spherical, metal-poor halo of our galaxy formed early during
the collapse of the proto-galactic cloud.  The collision at the
mid-plane halted the gas collapse and lead to the formation of the
rotating, thin disk.

However, there are many unanswered questions regarding the formation
of the Galaxy.  When did the bulge form? How and when did the thick
disk form?  How important is later infall and accretion?  Did the halo
form over an extended period of time?  Was halo formation a chaotic or
smooth process?  An important step towards answering these questions
is to determine accurate ages for the various stellar populations.
Globular clusters (GCs) play a key role in this regard, for their
derived ages are the most accurate of any object in the halo and thick
disk/bulge.  In this paper, ages for 43 Galactic GCs which have well
observed colour magnitude diagrams (CMDs) are derived and analyzed
to probe the formation of the Galactic halo.

Information regarding the formation and evolution of the Galaxy has
traditionally been obtained by surveys of stars or star clusters which
have one or more of the following properties measured: locations,
metallicities, velocities, and ages.  The classic paper by Eggen,
Lynden-Bell \& Sandage (1962) analyzed ultraviolet excesses, radial
velocities and proper motions of nearby stars to conclude that the
Galactic halo formed during the rapid, monolithic collapse of the
proto-Galactic gas cloud.  Evidence for a quite different halo
formation theory was presented by Searle \& Zinn (1978).  On the basis
of their studies of GC metallicities and horizontal branch morphology
Searle \& Zinn (1978) proposed that the halo formed via accretion
over several gigayears (Gyr) in a rather chaotic manner. These contrasting
theories continue to be a critical part of discussions of Galactic
formation (see Majewski 1993 for a recent review).  In this regard,
the determination of accurate absolute and relative ages for GCs plays an
important role in discovering the time scale for the formation of the
Galactic halo.

There are a variety of different methods which can be used to derive
the ages of GCs.  All of these techniques rely on comparing some
aspect of an observed CMD to theoretical stellar models or isochrones.
The most accurate relative ages can be derived using the difference in
colour between the main sequence turn-off and the base of the red
giant branch (\dbv, Sarajedini \& Demarque 1990; VandenBerg, Bolte \&
Stetson 1990).  However, the colours of theoretical isochrones are
very uncertain, as they depend on stellar atmospheres and the mixing
length treatment of convection.  As such, transforming an observed
difference in \dbv ~into an age difference is subject to large
theoretical uncertainties.  These uncertainties can be minimized by
comparing clusters with similar metallicities.

If one wishes to inter-compare ages of clusters with different
metallicities, then the difference in magnitude between the main
sequence turn-off and the horizontal branch (\dv) yields ages which
have the smallest theoretical errors.  Unfortunately, it is
often difficult to determine \dv ~observationally, so that the error
in the derived age can be rather large ($\sim 10\%$).  The absolute magnitude
of the main sequence turn-off (\mvto) is a well determined theoretical
quantity.  The new set of Yale isochrones (Chaboyer \ea 1995) provide
an up-to-date calibration of \mvto ~for a wide range of ages and
chemical compositions.  The absolute magnitude of the horizontal
branch (HB) is independent of age (over the range $\sim 8\sim 22$
Gyr), however its absolute level is not well determined in
theoretical models due to the importance of convection and
semi-convection in the nuclear burning regions of these stars.
Fortunately, there are a variety of independent, observationally based
methods which can be used to determine the absolute magnitude of RR
Lyr stars (\mvrr) which lie on the HB.  Hence, the \dv ~ages derived
in this paper are based on the calibration of \mvto ~as a function of age
and metallicity from the new set of Yale isochrones, coupled with a
variety of determinations of \mvrr.

There have been a number of studies of  \dv ~ages of GCs in recent
years (Sarajedini \& King 1989; Sandage \& Cacciari 1990; Carney,
Storm \& Jones 1992; Chaboyer, Sarajedini \& Demarque 1992; Walker
1992a; Caputo \ea 1993; Sandage 1993). These studies have shown that
the choice of a \mvrr ~relation is crucial to the conclusions which
are drawn based on \dv ~ages.  This work differs from previous studies
in three main ways: (i) use of the new Yale isochrones to determine
\mvto; (ii) an expanded observational database of observed \dv ~values
(30\% more than in our 1992 compilation); and (iii) the use of a large
number of \mvrr ~relations to explore in detail how the choice of
\mvrr ~affects our conclusions.

A brief description of the new Yale isochrones along with a discussion
of \mvrr ~and the theoretical calibration of \dv ~is presented in \S
2.  Section 3 reviews the basic observational data and tabulates the
\dv ~ages. A discussion of the correlations between age, metallicity
and galactocentric distance is contained in \S 4.  Evidence for an age
range within the Galactic globular cluster system is presented in \S
5.  Section 6 examines the second parameter problem in the context of
the \dv ~ages.  Finally, \S 7 discusses the major results of this
paper, and their implications for the formation of the Galactic halo.

\section{Theoretical Calibration of \dv}
\subsection{\mvto}
The recent set of Yale isochrones (Chaboyer \ea 1995) are used to
provide a calibration of \mvto ~as a function of age and metal
abundance.  These iso\-chrones are based on new stellar evolution models
which incorporate the latest available input physics:
opacities from Iglesias \& Rogers (1991, high temperature) and Kurucz
(1991, low temperature) and nuclear reaction rates from Bahcall \&
Pinsonneault (1992) and Bahcall (1989).  The colour transformation of
Green, Demarque \& King (1987) was used to construct isochrones in the
observational plane.  The new set of Yale isochrones are based on
standard stellar models, which do not include the effects of
diffusion, or the \dh ~correction to the equation of state.  Including
these two effects would systematically reduce the GC ages presented in
Table 3 by $\sim 13\%$ (Chaboyer 1995).

In order to span the range of metallicities of observed globular
clusters, \mvto ~values were determined from isochrones with $\feh =
-2.8,\,-2.3,$ $-1.8, \,-1.3,\,-1.0$ and $-0.44$.  The isochrones with
$\feh \le -1.0$ have a helium abundance of $Y=0.23$. This is in good
agreement with recent determinations of the primordial helium
abundance (Pagel \& Kazlauskas 1992; Balbes, Boyd \& Mathews 1993;
Izotov, Thuan \& Lipovetsky 1994).  The most metal rich isochrone
($\feh = -0.44$) has a helium abundance of $Y= 0.25$ and assumes that
the $\alpha$-capture elements (O, Mg, Si, S, and Ca) are
enhanced by 0.2 dex over their solar values (ie.~$[\alpha/{\rm Fe}] =
+0.2$).  The more metal-poor isochrones assume $[\alpha/{\rm Fe}] =
+0.4$.  These $[\alpha/{\rm Fe}]$ values were chosen to be in
agreement with observations of halo stars (Lambert 1989; Dickens \ea
1991; King 1994; Nissen \ea 1994). The new Yale isochrones are
tabulated every 2 Gyr for the older ages (10 -- 22 Gyr). In order to
provide a finer grid for interpolation purposes in this study, we have
recomputed these isochrones using a 1 Gyr spacing, between 8 -- 22
Gyr.

\subsection{\mvrr}
There are numerous observational and theoretical techniques which may
be used to derive \mvrr.  There is general agreement that the absolute
magnitude of RR Lyr stars is given by an equation of the form
\beq
\mvrr = \mu\, \feh + \gamma,
\label{rrlyr}
\eeq where $\mu$ is the slope with metallicity and $\gamma$ is the
zero-point.  The zero-point is important for setting the overall
absolute ages, while the slope is important in determining the
relative ages for GCs with different metallicities.  Some techniques
for determining \mvrr ~are best for determining the zero-point, while
other techniques are best at deriving the slope.  Due to possible
systematic effects, the Baade-Wesselink and infrared flux analysis
are best used to determine the slope with metallicity.  Such
analyses of field RR Lyr stars have been published recently by two
groups.  Jones \ea (1992) found $\mu = 0.16\pm 0.03$, while Skillen
\ea (1993) determined $\mu = 0.21\pm 0.05$.  The theoretical HB models
of Lee (1990) should also give a reliable determination of the slope,
and yield $\mu = 0.18\pm 0.01$.  From an analysis of the Oosterhoff
period shift effect in GCs, Sandage (1993)\footnote{Sandage (1993) gives
an error of 0.12 for the slope of $\rm
M_{bol}\,(RR)$ with metallicity, but does not quote an error in \mvrr.
However, Sandage obtains \mvrr ~from a simple transformation of his
$\rm M_{bol}\,(RR)$ equation (${\rm M_{bol}} = {\rm M_V} + 0.06\,\feh
+ 0.06$), so we have taken the error in his \mvrr ~relation from his
quoted error in his $\rm M_{bol}\,(RR)$ relation.}
found $\mu = 0.30\pm 0.12$.  Thus, it
appears that \mvrr ~has a rather shallow slope with metallicity of
$\mu \simeq 0.20$, though it may be somewhat premature to totally
exclude slopes as high as $\mu = 0.30$.

A reliable determination of the zero-point in equation (\ref{rrlyr})
can be made by measuring the apparent magnitude of a number of RR
Lyr stars in the LMC and then using the distance to the LMC to
obtain $\gamma$.  This is the approach used by Walker (1992a) who
found $\mvrr = 0.44\pm 0.10$ at $\feh = -1.9$ (implying $\gamma =
0.82\pm 0.10$), assuming $({\rm m-M})_{\rm LMC} = 18.5\pm 0.10$.  This
distance modulus to the LMC was based on main sequence fitting, and
analysis of the Cepheid variables and the rings associated with
SN1987A.  However, Gould (1995) has recently re-analyzed the SN1987A
distance estimate, and determined an upper limit of $({\rm m-M})_{\rm
LMC} = 18.37$.  Using this distance estimate to the LMC and Walker's
(1992a) RR Lyr photometry, one finds $\mvrr = 0.57$ at $\feh = -1.9$
($\gamma = 0.95$).  Using a statistical parallax analysis of field RR
Lyr stars, Layden, Hanson \& Hawley (1994) found a zero-point which is
0.24 mag fainter ($\gamma = 1.06$) than that quoted by Walker (1992a)
(and 0.11 mag fainter than the revised Walker value above).  This
suggests that $\gamma \simeq 0.98$ is a reasonable choice, and so our
best estimate for the absolute magnitude of the RR Lyr stars is $\mvrr
= 0.20\feh + 0.98$.  This choice of \mvrr ~predicts that the RR Lyrs
in M92 (with $\feh = -2.24\pm 0.08$) should have $\mvrr_{\rm M92} =
0.53\pm 0.02$.  In their Baade-Wesselink analysis of M92 RR Lyrs,
Storm, Carney \& Latham (1994) determined $\mvrr_{\rm M92} = 0.43\pm
0.22$, which is in reasonable agreement.  However, the possible 0.1
mag discrepancy would reduce the ages for the metal-poor clusters by
10\% and so a more precise determination of \mvrr ~in metal-poor
GCs is clearly desirable.  Although we have derived our best estimate
for the absolute magnitude of the RR Lyr stars ($\mvrr = 0.20\feh +
0.98$), in order to explore in a systematic way the effect that
uncertainties in \mvrr ~have on the GC ages, ages will be derived
using both the Walker (1992a) and Layden \ea (1994) zero-points with
slopes which vary from 0.15 to 0.30.
\footnotesize
\begin{planotable}{lrrrrrrrrrrrrr}
\tablewidth{0pt}
\tablecaption{TABLE 1}
\tablecaption{Fitting Coefficients}
\tablehead{
& \multicolumn{6}{c}{Standard} &&
\multicolumn{6}{c}{Red HB} \nl
\cline{2-7}\cline{9-14}
{}~\\[-8pt]
\colhead{\mvrr} &
\colhead{$\beta_0$} &
\colhead{$\beta_1$} &
\colhead{$\beta_2$} &
\colhead{$\beta_3$} &
\colhead{$\beta_4$} &
\colhead{$\beta_5$} & &
\colhead{$\beta_0$} &
\colhead{$\beta_1$} &
\colhead{$\beta_2$} &
\colhead{$\beta_3$} &
\colhead{$\beta_4$} &
\colhead{$\beta_5$}
}
\startdata
$0.17 \,\feh + 0.79$ &
$ 74.995$ &  $-47.943$ &  $  8.435$ &  $  6.219$ &  $0.551$ &$-2.025$ &&
$100.913$ &  $-62.728$ &  $ 10.550$ &  $ 21.501$ &  $3.562$ &$-5.436$\nl
$0.20 \,\feh + 0.98$ &
$ 77.703$ &  $-51.506$ &  $  9.419$ &  $  4.669$ &  $0.457$ &$-1.611$ &&
$ 88.669$ &  $-58.294$ &  $ 10.485$ &  $ 18.692$ &  $3.385$ &$-4.810$\nl
$0.15 \,\feh + 0.98$ &
$ 83.026$ &  $-54.407$ &  $  9.813$ &  $  7.659$ &  $0.573$ &$-2.664$ &&
$ 81.215$ &  $-53.656$ &  $  9.768$ &  $ 22.377$ &  $3.664$ &$-6.078$ \nl
$0.15 \,\feh + 0.72$ &
$ 76.664$ &  $-48.252$ &  $  8.335$ &  $  7.177$ &  $0.614$ &$-2.297$ &&
$105.035$ &  $-64.100$ &  $ 10.550$ &  $ 23.136$ &  $3.675$ &$-5.858$ \nl
$0.20 \,\feh + 1.06$ &
$ 78.270$ &  $-52.404$ &  $  9.725$ &  $  5.012$ &  $0.456$ &$-1.752$ &&
$ 69.791$ &  $-47.837$ &  $  9.144$ &  $ 19.122$ &  $3.474$ &$-5.006$ \nl
$0.20 \,\feh + 0.82$ &
$ 72.434$ &  $-46.790$ &  $  8.343$ &  $  4.684$ &  $0.503$ &$-1.499$ &&
$ 99.041$ &  $-62.095$ &  $ 10.550$ &  $ 19.475$ &  $3.408$ &$-4.803$ \nl
$0.25 \,\feh + 1.14$ &
$ 74.978$ &  $-51.177$ &  $  9.753$ &  $  2.728$ &  $0.406$ &$-0.880$ &&
$ 85.981$ &  $-57.814$ &  $ 10.783$ &  $ 19.077$ &  $3.341$ &$-4.855$ \nl
$0.25 \,\feh + 0.91$ &
$ 67.189$ &  $-44.791$ &  $  8.296$ &  $  2.212$ &  $0.434$ &$-0.656$ &&
$ 93.237$ &  $-60.091$ &  $ 10.550$ &  $ 16.014$ &  $3.195$ &$-3.748$ \nl
$0.30 \,\feh + 1.22$ &
$ 72.167$ &  $-50.378$ &  $  9.842$ &  $  0.000$ &  $0.240$ &$ 0.000$ &&
$ 81.425$ &  $-56.089$ &  $ 10.783$ &  $ 15.884$ &  $3.125$ &$-3.776$ \nl
$0.30 \,\feh + 1.01$ &
$ 67.561$ &  $-45.831$ &  $  8.638$ &  $  0.000$ &  $0.249$ &$ 0.000$ &&
$ 86.174$ &  $-57.216$ &  $ 10.420$ &  $ 12.826$ &  $3.024$ &$-2.718$ \nl
\label{tab1}
\end{planotable}
\normalsize

\subsection{Derivation of \dv ~Ages}
The \mvto ~values from the new Yale isochrones are combined with a
given \mvrr ~relation to form a grid, which specifies age given \dv
{}~and \feh.  The grid is then fit to an equation of the form
\beqa
t_9 &=& \beta_0 + \beta_1\dvtwo + \beta_2\dvtwo^2 + \beta_3\feh
\nonumber \\
&+ &\beta_4\feh^2 + \beta_5\dvtwo\feh,
\label{fit}
\eeqa
where $t_9$ is the age in Gyr.  The rms residuals of the points from
the fit were about 0.15 Gyr.  The above formula is used to determine
ages for GC which have RR Lyrs, or a blue HB.  In the case clusters
with purely blue HBs (i.e. few or no RR Lyrae variables), observers
quote the V mag of the blue edge of the instability strip, which is
usually a reasonably accurate measurement of \mvrr.  In some cases,
they compare to the blue HB of a cluster with RR Lyraes to infer
\mvrr.

In the case of clusters with red HBs (HB type\footnote{The HB type has the
following definition: $\rm HB\,type \equiv (B - R)/(B+V+R)$, where B
is the number of HB stars blueward of the instability strip, R is the
number of HB stars redward of the instability strip, and V is the
number of RR Lyr stars} $\le -0.8$.), the situation is slightly
more complicated.  In these clusters, observers usually quote the
mean or median mag of the red HB stars.  This quantity can be anywhere from
0.05 to 0.2 mag brighter or fainter than the RR Lyr level depending on
the cluster metallicity and age.  In order to correct for this effect,
a semi-empirical approach is taken.  The offset between the red HB
level and RR Lyr level may be determined from theoretical HB models,
and this correction can then be applied to the red HB clusters, as
discussed by Fullton \ea (1995).  As this offset depends on relative
quantities in the theoretical models, it should be reasonably
reliable.  HB models by Lee, Demarque \& Zinn (1987), Dorman (1992)
and Castellani, Chieffi \& Pulone (1991) find offsets which agree to
within 0.05 mag.  The offsets used in this study are derived from HB
models kindly provided to us by Lee (1995).  Lee constructed synthetic
HB models of red HB clusters with a range of ages and abundance from
which he has calculated $\rm M_V(HB)$.  Given the relation $\mvrr =
0.17\,\feh + 0.79$, which comes from the Lee HB models, the function
$\delta = {\rm M_V(red\,HB)} - \mvrr$ can be calculated as a
function of \feh ~and age. For other RR Lyrae luminosity relations,
the function $\delta$ is used to correct \mvrr ~to $\rm M_V(red\,HB)$.

The coefficients of the fit (equation \ref{fit}) for the standard and
red HB cases are given in Table 1.  These coefficients have been
tabulated for 10 different \mvrr ~relations: the Lee (1995) relation
($\mvrr = 0.17\,\feh + 0.79$); our best estimate for the true relation
($\mvrr = 0.20\,\feh + 0.98$); and relations with slopes of 0.15,
0.20, 0.25 and 0.30 using the zero-points of Walker (1992a) and Layden
\ea (1994).  These \mvrr ~relations span the range reported by various
groups using a variety of observational and theoretical techniques
(see \S 2.2).  The coefficients presented in Table 1 are valid for
$-2.8\le \feh \le -0.44$ and for ages in the range $8 - 22$
Gyr.

\section{The Ages}
Estimating GC ages using \dv ~requires an accurate measurement of
V(TO) and V(HB), along with an estimate of \feh.  Continued advances
in CCD technology and image reduction, along with the advent of the
HST have lead to a wealth of high quality CMDs of GCs in recent years.
Table 2 lists various observational quantities for 43 GCs for which
reliable age determinations may be made based on published
observations of V(TO) and V(HB). To be conservative, M22 and
$\omega\,$Cen have not been included in this group, as there is
evidence for a range in metallicity in these clusters (Noble \ea 1991;
Smith 1987) which complicates the age determination process.
References for V(TO) and V(HB) are provided in the table.  In some
cases, the observers do not quote V(HB); rather, they provide the
apparent magnitude of the zero-age horizontal branch (V(ZAHB)).  These
have been converted to mean HB magnitudes using equation 4 from Carney
\ea (1992: $\rm V(HB) = V(ZAHB) - 0.05\feh -0.20$).  The V(HB) (and
corresponding \dv ~values) which have been corrected for this effect
are indicated by an asterisk next to the V(HB) value in Table 2.  Our
1992 compilation (Chaboyer \ea 1992) does NOT include a correction for
this effect, as we were not aware that some observers quoted V(ZAHB).
\footnotesize
\begin{planotable}{rlrrrrllrrrrcc}
\tablewidth{0pt}
\tablecaption{TABLE 2}
\tablecaption{Globular Cluster Parameters}
\tablehead{
\multicolumn{2}{c}{Cluster}&
\colhead{$\ell$}&
\colhead{b}& &&&&
\colhead{$\rm R_{GC}$}&
\colhead{$\rm R_{apo}$}& & &
\multicolumn{2}{c}{References\tablenotemark{d}}\\[.2ex]
\colhead{NGC}&
\colhead{Name}&
\colhead{(degrees)}&
\colhead{(degrees)}&
\colhead{~$\rm E(B-V)$}&
\colhead{\feh}&
\colhead{V(HB)\tablenotemark{a}}&
\colhead{\dv\tablenotemark{b}}&
\colhead{(kpc)}&
\colhead{(kpc)}&
\colhead{HB type} &
\colhead{Group\tablenotemark{c}} &
\colhead{~HB}&
\colhead{TO}
}
\startdata
 104& 47 Tuc& $305.896$& $-44.900$& $0.04$& $-0.71 \pm 0.08$& $14.09$& $3.61
\pm 0.10$& $ 7.3$& $ 7.3$& $-1.00 \pm 0.03$&  D&  1&  1 \nl
 288&       & $151.328$& $-89.383$& $0.04$& $-1.40 \pm 0.12$& $15.27^*$& $3.73
\pm 0.12$& $11.2$& $11.5$& $ 0.95 \pm 0.08$& OH&  2&  2 \nl
 362&       & $301.533$& $-46.248$& $0.06$& $-1.27 \pm 0.07$& $15.29^*$& $3.56
\pm 0.14$& $ 8.8$& $ 9.2$& $-0.87 \pm 0.08$& YH&  2&  2 \nl
1261&       & $270.541$& $-52.126$& $0.00$& $-1.31 \pm 0.09$& $16.57^*$& $3.57
\pm 0.12$& $16.8$& \multicolumn{1}{c}{---}& $-0.70 \pm 0.10$& YH&  3&  3 \nl
1851&       & $244.514$& $-35.037$& $0.02$& $-1.36 \pm 0.09$& $16.15$& $3.45
\pm 0.10$& $16.5$& \multicolumn{1}{c}{---}& $-0.33 \pm 0.08$& YH&  4&  4 \nl
1904& M79   & $227.229$& $-29.351$& $0.01$& $-1.69 \pm 0.09$& $16.03^*$& $3.57
\,\pm\;${\bf ---}& $17.8$& \multicolumn{1}{c}{---}& $ 0.89 \pm 0.16$& OH&  5&
5 \nl
2298&       & $245.628$& $-16.007$& $0.08$& $-1.85 \pm 0.11$& $16.11$& $3.49
\pm 0.21$& $16.1$& \multicolumn{1}{c}{---}& $ 0.93 \pm 0.24$& OH&  6&  6 \nl
2808&       & $282.192$& $-11.253$& $0.22$& $-1.37 \pm 0.09$& $15.97^*$& $3.63
\pm 0.14$& $10.2$& \multicolumn{1}{c}{---}& $-0.54 \pm 0.06$& YH&  2&  2 \nl
3201&       & $277.228$& $  8.641$& $0.21$& $-1.61 \pm 0.12$& $14.76$& $3.39
\pm 0.17$& $ 8.8$& \multicolumn{1}{c}{---}& $ 0.08 \pm 0.06$& YH&  7&  8 \nl
4147&       & $252.850$& $ 77.189$& $0.02$& $-1.80 \pm 0.14$& $17.00$& $3.60
\,\pm\;${\bf ---}& $20.5$& $25.7$& $ 0.55 \pm 0.14$& YH&  9&  9 \nl
4590& M68   & $299.626$& $ 36.052$& $0.07$& $-2.09 \pm 0.11$& $15.64$& $3.42
\pm 0.10$& $ 9.6$& \multicolumn{1}{c}{---}& $ 0.44 \pm 0.05$& YH& 10& 10 \nl
5024& M53   & $332.965$& $ 79.765$& $0.00$& $-2.04 \pm 0.08$& $16.90$& $3.55
\pm 0.14$& $18.9$& \multicolumn{1}{c}{---}& $ 0.76 \pm 0.10$& OH& 11& 12 \nl
5053&       & $335.695$& $ 78.944$& $0.06$& $-2.41 \pm 0.06$& $16.65$& $3.48
\,\pm\;${\bf ---}& $16.2$& \multicolumn{1}{c}{---}& $ 0.61 \pm 0.18$& OH& 13&
14 \nl
5272& M3    & $ 42.208$& $ 78.708$& $0.01$& $-1.66 \pm 0.06$& $15.63$& $3.54
\pm 0.09$& $11.7$& $18.3$& $ 0.08 \pm 0.04$& YH& 15& 15 \nl
5466&       & $ 42.150$& $ 73.592$& $0.00$& $-2.22 \pm 0.36$& $16.62$& $3.58
\,\pm\;${\bf ---}& $16.7$& $29.2$& $ 0.68 \pm 0.14$& OH& 16& 17 \nl
5897&       & $342.946$& $ 30.294$& $0.06$& $-1.68 \pm 0.11$& $16.35$& $3.60
\pm 0.18$& $ 7.6$& \multicolumn{1}{c}{---}& $ 0.91 \pm 0.10$& OH& 18& 18 \nl
5904& M5    & $  3.863$& $ 46.796$& $0.03$& $-1.40 \pm 0.06$& $14.98^*$& $3.62
\pm 0.11$& $ 6.0$& $21.1$& $ 0.37 \pm 0.06$& OH&  2&  2 \nl
6101&       & $317.747$& $-15.825$& $0.04$& $-1.81 \pm 0.15$& $16.60$& $3.40
\,\pm\;${\bf ---}& $10.7$& \multicolumn{1}{c}{---}& $ 0.84 \pm 0.16$& OH& 19&
19 \nl
6121& M4    & $350.974$& $ 15.972$& $0.40$& $-1.33 \pm 0.10$& $13.22^*$& $3.68
\pm 0.16$& $ 6.4$& $ 6.4$& $-0.07 \pm 0.10$& OH&  2&  2 \nl
6171& M107  & $  3.374$& $ 23.011$& $0.31$& $-0.99 \pm 0.06$& $15.55^*$& $3.75
\pm 0.18$& $ 3.6$& $ 3.9$& $-0.76 \pm 0.08$& OH&  2&  2 \nl
6205& M13   & $ 59.008$& $ 40.912$& $0.02$& $-1.65 \pm 0.06$& $14.83^*$& $3.67
\pm 0.21$& $ 8.2$& $18.9$& $ 0.97 \pm 0.08$& OH&  2&  2 \nl
6218& M12   & $ 15.717$& $ 26.313$& $0.17$& $-1.34 \pm 0.09$& $14.90$& $3.45
\,\pm\;${\bf ---}& $ 4.3$& $ 4.5$& $ 0.92 \pm 0.10$& OH& 20& 21 \nl
6254& M10   & $ 15.138$& $ 23.077$& $0.32$& $-1.60 \pm 0.08$& $14.65$& $3.75
\pm 0.15$& $ 4.9$& \multicolumn{1}{c}{---}& $ 0.94 \pm 0.10$& OH& 22& 23 \nl
6341& M92   & $ 68.339$& $ 34.860$& $0.02$& $-2.24 \pm 0.08$& $14.96^*$& $3.74
\pm 0.12$& $ 9.1$& $ 9.3$& $ 0.88 \pm 0.08$& OH&  2&  2 \nl
6352&       & $341.421$& $ -7.167$& $0.21$& $-0.51 \pm 0.08$& $15.13$& $3.67
\pm 0.10$& $ 3.6$& \multicolumn{1}{c}{---}& $-1.00 \pm 0.04$&  D&  1&  1 \nl
6397&       & $338.165$& $-11.958$& $0.18$& $-1.91 \pm 0.14$& $12.90^*$& $3.74
\pm 0.14$& $ 6.1$& $ 6.1$& $ 0.93 \pm 0.10$& OH&  2&  2 \nl
6535&       & $ 27.177$& $ 10.436$& $0.44$& $-1.75 \pm 0.15$& $15.73$& $3.66
\pm 0.19$& $ 4.2$& \multicolumn{1}{c}{---}& $ 1.00 \,\pm\;${\bf
---}\hspace*{5pt}& OH& 24& 24 \nl
6584&       & $342.144$& $-16.414$& $0.07$& $-1.54 \pm 0.15$& $16.53$& $3.47
\,\pm\;${\bf ---}& $ 6.9$& \multicolumn{1}{c}{---}& $-0.09 \pm 0.06$& YH& 25&
25 \nl
6652&       & $  1.534$& $-11.377$& $0.10$& $-0.89 \pm 0.15$& $15.85$& $3.35
\pm 0.16$& $ 1.9$& \multicolumn{1}{c}{---}& $-1.00 \,\pm\;${\bf
---}\hspace*{5pt}& OH& 26& 26 \nl
6752&       & $336.496$& $-25.627$& $0.04$& $-1.54 \pm 0.09$& $13.63^*$& $3.77
\pm 0.16$& $ 5.4$& \multicolumn{1}{c}{---}& $ 1.00 \pm 0.04$& OH&  2&  2 \nl
6809& M55   & $  8.795$& $-23.270$& $0.06$& $-1.82 \pm 0.15$& $14.24^*$& $3.66
\pm 0.10$& $ 4.1$& \multicolumn{1}{c}{---}& $ 0.91 \pm 0.10$& OH&  2&  2 \nl
6838& M71   & $ 56.744$& $ -4.564$& $0.27$& $-0.58 \pm 0.08$& $14.44$& $3.56
\pm 0.09$& $ 6.8$& $ 6.7$& $-1.00 \pm 0.04$&  D& 27& 27 \nl
7006&       & $ 63.770$& $-19.407$& $0.05$& $-1.59 \pm 0.07$& $18.80$& $3.55
\pm 0.12$& $36.8$& \multicolumn{1}{c}{---}& $-0.11 \pm 0.06$& YH& 28& 28 \nl
7078& M15   & $ 65.013$& $-27.313$& $0.10$& $-2.15 \pm 0.08$& $15.77^*$& $3.63
\pm 0.16$& $ 9.9$& $10.2$& $ 0.72 \pm 0.10$& OH&  2&  2 \nl
7099& M30   & $ 27.180$& $-46.835$& $0.04$& $-2.13 \pm 0.13$& $15.11^*$& $3.62
\pm 0.14$& $ 6.9$& \multicolumn{1}{c}{---}& $ 0.88 \pm 0.12$& OH&  2&  2 \nl
7492&       & $ 53.386$& $-63.478$& $0.00$& $-1.82 \pm 0.30$& $17.52^*$& $3.72
\pm 0.14$& $23.2$& \multicolumn{1}{c}{---}& $ 0.90 \pm 0.18$& OH&  2&  2 \nl
   & Ter 7  & $  3.387$& $-20.063$& $0.06$& $-0.36 \pm 0.09$& $17.76$& $3.20
\pm 0.12$& $14.3$& \multicolumn{1}{c}{---}& $-1.00 \,\pm\;${\bf
---}\hspace*{5pt}& YH& 29& 29 \nl
   & Ter 8  & $  5.758$& $-24.558$& $0.20$& $-1.99 \pm 0.08$& $17.85$& $3.65
\,\pm\;${\bf ---}& $14.3$& \multicolumn{1}{c}{---}& $ 1.00 \,\pm\;${\bf
---}\hspace*{5pt}& OH& 30& 31 \nl
   & Rup 106& $300.888$& $ 11.670$& $0.20$& $-1.69 \pm 0.05$& $17.73^*$& $3.32
\pm 0.07$& $17.0$& \multicolumn{1}{c}{---}& $-0.82 \pm 0.15$& YH& 32& 32 \nl
   & Pal 5  & $  0.852$& $ 45.860$& $0.03$& $-1.47 \pm 0.29$& $17.27^*$& $3.53
\pm 0.14$& $15.4$& $16.9$& $-0.40 \pm 0.20$& YH&  2&  2 \nl
   & Pal 12 & $ 30.510$& $-47.680$& $0.02$& $-1.14 \pm 0.20$& $17.13$& $3.30
\,\pm\;${\bf ---}& $15.1$& \multicolumn{1}{c}{---}& $-1.00 \pm 0.12$& YH& 33&
33 \nl
   & IC4499& $307.354$& $-20.473$& $0.25$& $-1.75 \pm 0.20$& $17.80$& $3.25 \pm
0.15$& $15.7$& \multicolumn{1}{c}{---}& $ 0.11 \pm 0.36$& YH& 34& 34 \nl
   & Arp 2  & $  8.543$& $-20.787$& $0.08$& $-1.70 \pm 0.11$& $18.18^*$& $3.40
\pm 0.10$& $21.5$& \multicolumn{1}{c}{---}& $ 0.86 \,\pm\;${\bf
---}\hspace*{5pt}& OH& 35& 35 \nl

\tablenotetext{a}{Entries with an asterick ($^*$) were origininally
published as V(ZAHB).  The V(HB) and \dv ~values have been corrected
using $\rm V(HB) = V_{ZAHB} - 0.05\feh - 0.20$ (eqn.~4 from Carney \ea 1992).}

\tablenotetext{b}{Errors given in \dv ~are those quoted in the
original papers, and are not $1\,\sigma$ Gaussian errors.
As discussed in Appendix A, a reasonable estimate of the $1\,\sigma$
Gaussian error may be obtained by multipling the quoted errors by 0.61}

\tablenotetext{c}{D$\,\equiv\,$Disk; OH$\,\equiv\,$Old Halo;
YH$\,\equiv\,$Younger Halo}

\tablenotetext{d}{(1) Fullton \ea 1995;
(2) Buonanno, Coris \& Fusi Pecci 1989; (3) Ferraro \ea 1993; (4)
Walker 1992b; (5) Ferraro \ea 1992a; (6) Janes \& Heasley 1988; (7)
Alcaino, Liller \& Alvarado 1989; (8) Brewer \ea 1993; (9) Friel,
Heasley \& Christian 1987; (10) Walker 1994; (11) Cuffey 1965; (12)
Heasley \& Christian 1991; (13) Sarajedini \& Milone 1995; (14)
Fahlman, Richer \& Nemec 1991; (15) Buonanno \ea 1994; (16) Nemec \&
Harris 1987; (17) Peterson 1986; (18) Sarajedini 1992; (19) Sarajedini
\& Da Costa 1991; (20) Racine 1971; (21) Sato, Richer \& Fahlman 1989;
(22) Hurley, Richer \& Fahlman 1989; (23) Harris, Racine \& deRoux
1976; (24) Sarajedini 1994; (25) Sarajedini \& Forrester 1995; (26)
Ortolani, Bica \& Barbuy 1994; (27) Hodder \ea 1992; (28) Buonanno \ea
1991; (29) Buonanno \ea 1995b; (30) Da Costa \& Armandroff 1995; (31)
Ortolani \& Gratton 1990; (32) Buonanno \ea 1993; (33) Stetson \ea
1989; (34) Ferraro \ea (1995); (35) Buonanno \ea 1995a}
\label{tab2}
\end{planotable}
\normalsize

The cluster \feh ~values and their errors are taken from Zinn \& West
(1984), except for NGC 4147, 5053 (Armandroff, Da Costa \& Zinn 1992),
NGC 6218, Ter 7, Ter 8, and Arp 2 (Da Costa \& Armandroff 1995), and
Rup 106 (Da Costa, Armandroff, \& Norris 1992).  The reddennings are
from Zinn (1985), except for NGC 4590 (Walker 1994), NGC 5053
(Sarajedini \& Milone 1995), NGC 6352 (Fullton \ea 1995), NGC 6535
(Sarajedini 1994), NGC 6584 (Sarajedini \& Forrester 1995), NGC 6652
(Ortolani, Bica \& Barbuy 1994), Ter 7 (Webbink 1985), Ter 8 (Ortolani
\& Gratton 1990), Arp 2 (Buonanno \ea (1995a) and Rup 106 (Da Costa,
Armandroff \& Norris 1992).  The Galactic coordinates for the clusters
are taken from Shawl \& White (1986), except for Rup 106 and Pal 12,
which are from Webbink (1985).  The HB types are from Lee, Demarque \&
Zinn (1994), with the exceptions of NGC 4590 (Walker 1994), NGC 6584
(Sarajedini \& Forrester 1995), NGC 6535 (Sarajedini 1994), NGC 6652
(Ortolani \ea 1994), Ter 7, Arp 2 (Buonanno \ea 1995a,b), Ter 8
(Ortolani \& Gratton 1990), and IC4499 (Ferraro \ea 1995).  The
groupings into disk, old halo and younger halo clusters are from Zinn
\& Lee (1995; see also Zinn 1993).  In calculating the Galactocentric
distance ($\rm R_{GC}$) of each cluster, we have adopted $R_{\odot} =
8.0\,$kpc, $\rm A_V = 3.2E(B-V)$, and a distance modulus derived from
V(HB) and our preferred \mvrr relation ($\mvrr = 0.20\,\feh + 0.98$).
Proper motion studies exist for 16 of the GCs in our sample, and these
have been used by Majewski (1994) to determine the apogalactica
distances ($\rm R_{apo}$) listed in Table 2.  As expected, most of the
apogalactica distances are quite similar to the Galacto-centric
distances.  However, there are a few notable exceptions.  The present
positions of NGC 5466, 5904 and 6205 are considerably smaller than
their apogalactica distances.

Using the \dv ~and \feh ~values listed in Table 2, GC ages are derived
using equation (\ref{fit}), with the $\beta$ coefficients listed in
Table 1.  The error in the derived age is calculated by propagating
the errors in \dv ~and \feh ~through equation (\ref{fit}).  The error
in the derived age is dominated by the error in \dv.  For the
statistical analysis which comprises the bulk of this paper, it is
important that the error in \dv ~represent a Gaussian 1-sigma error bar.
However, it is doubtful that the observers quote such an error bar.
In order to get an estimate for the Gaussian 1-sigma error bar, the
literature has been searched for independent measurements of \dv.
Appendix A presents an analysis of these independent observations, and
concludes that a reliable estimate for the Gaussian 1-sigma error in
\dv ~may be obtained by multiplying the quoted error by 0.61.  This
correction factor has been applied to all of the quoted \dv ~errors
when determining the error in the derived age.  In some cases, errors
in \dv ~were not given by the observers.  For these clusters, an error
of 0.083 mag was assumed.  This is the average of the Gaussian
1-sigma errors for those clusters with quoted errors in \dv.

Table 3 presents ages for the 43 GCs, using the 10 different \mvrr
{}~relations given in Table 1.  The heading of each column gives the
\mvrr ~relation used to derive the ages in that column.  These ages
will be analyzed in detail in the following sections.  Here, we simply
note that ages derived using the Layden \ea (1994) zero-point for
\mvrr ~are approximately 25\% larger than the ages derived using the
Walker zero-point (e.g.~compare columns 5 and 6).  This illustrates
the well known result that a 0.25 mag uncertainty in the distance
modulus translates into a 25\% uncertainty in ages derived using \dv.
It is also interesting to note that there are several young clusters
in the sample; IC4499, Arp 2, Pal 12, Rup 106, and Ter 7 have all been
shown to be young by the \dbv ~technique (see Buonanno \ea 1994) and
indeed, the \dv ~ages for these clusters are all young compared to the
mean age.  NGC 6652, which has a small \dv ~value (Ortolani, Bica \&
Barbuy 1994), also appears to be young.
\footnotesize
\begin{planotable}{lrrrrrrrrrr}
\tablewidth{0pt}
\tablecaption{TABLE 3}
\tablecaption{Globular Cluster Ages}
\tablehead{
&
\colhead{$0.17\,\feh$}&
\colhead{$0.20\,\feh$}&
\colhead{$0.15\,\feh$}&
\colhead{$0.15\,\feh$}&
\colhead{$0.20\,\feh$}&
\colhead{$0.20\,\feh$}&
\colhead{$0.25\,\feh$}&
\colhead{$0.25\,\feh$}&
\colhead{$0.30\,\feh$}&
\colhead{$0.30\,\feh$}\\[.2ex]
&
\colhead{$+ 0.79$}      &
\colhead{$+ 0.98$}      &
\colhead{$+ 0.98$}      &
\colhead{$+ 0.725$}      &
\colhead{$+ 1.06$}      &
\colhead{$+ 0.82$}      &
\colhead{$+ 1.14$}      &
\colhead{$+ 0.915$}      &
\colhead{$+ 1.22$} &
\colhead{$+ 1.01$}\\[.2ex]
Name&
\colhead{Age}&
\colhead{Age}&
\colhead{Age}&
\colhead{Age}&
\colhead{Age}&
\colhead{Age}&
\colhead{Age}&
\colhead{Age}&
\colhead{Age}&
\colhead{Age}
}
\startdata
{}~104& $12.4\pm 1.1 $&$15.6\pm 1.3 $&$16.4\pm 1.3 $&$11.6\pm 1.0 $&$17.3\pm
1.4 $&$12.6\pm 1.1 $&$18.4\pm 1.5 $&$13.6\pm 1.2 $&$19.4\pm 1.5 $&$14.8\pm 1.2
$ \nl
{}~288& $16.5\pm 1.3 $&$19.4\pm 1.6 $&$20.9\pm 1.7 $&$15.8\pm 1.3 $&$21.1\pm
1.7 $&$16.2\pm 1.3 $&$21.4\pm 1.7 $&$16.7\pm 1.3 $&$21.7\pm 1.7 $&$17.3\pm 1.4
$ \nl
{}~362& $14.3\pm 1.7 $&$17.4\pm 2.0 $&$18.9\pm 2.1 $&$13.5\pm 1.6 $&$19.3\pm
2.1 $&$14.1\pm 1.7 $&$19.8\pm 2.2 $&$14.7\pm 1.7 $&$20.3\pm 2.2 $&$15.3\pm 1.8
$ \nl
1261& $13.7\pm 1.1 $&$16.2\pm 1.3 $&$17.4\pm 1.4 $&$13.1\pm 1.1 $&$17.6\pm 1.4
$&$13.5\pm 1.1 $&$17.9\pm 1.4 $&$14.0\pm 1.1 $&$18.3\pm 1.5 $&$14.5\pm 1.2 $
\nl
1851& $12.0\pm 0.8 $&$14.2\pm 1.0 $&$15.3\pm 1.1 $&$11.6\pm 0.8 $&$15.5\pm 1.1
$&$11.9\pm 0.8 $&$15.7\pm 1.1 $&$12.3\pm 0.8 $&$16.0\pm 1.1 $&$12.7\pm 0.8 $
\nl
1904& $14.5\pm 1.3 $&$16.9\pm 1.5 $&$18.5\pm 1.7 $&$14.1\pm 1.3 $&$18.4\pm 1.7
$&$14.2\pm 1.3 $&$18.3\pm 1.7 $&$14.4\pm 1.3 $&$18.4\pm 1.6 $&$14.7\pm 1.3 $
\nl
2298& $13.9\pm 1.9 $&$16.0\pm 2.2 $&$17.7\pm 2.5 $&$13.4\pm 1.8 $&$17.4\pm 2.4
$&$13.5\pm 1.8 $&$17.2\pm 2.4 $&$13.5\pm 1.8 $&$17.1\pm 2.3 $&$13.7\pm 1.9 $
\nl
2808& $14.7\pm 1.4 $&$17.3\pm 1.6 $&$18.7\pm 1.8 $&$14.1\pm 1.3 $&$18.9\pm 1.8
$&$14.5\pm 1.4 $&$19.2\pm 1.8 $&$15.0\pm 1.4 $&$19.5\pm 1.8 $&$15.5\pm 1.4 $
\nl
3201& $11.9\pm 1.3 $&$13.8\pm 1.6 $&$15.0\pm 1.7 $&$11.5\pm 1.3 $&$15.1\pm 1.7
$&$11.6\pm 1.3 $&$15.0\pm 1.7 $&$11.8\pm 1.3 $&$15.1\pm 1.7 $&$12.1\pm 1.3 $
\nl
4147& $15.4\pm 1.4 $&$17.9\pm 1.6 $&$19.7\pm 1.8 $&$14.9\pm 1.4 $&$19.5\pm 1.8
$&$15.0\pm 1.4 $&$19.3\pm 1.7 $&$15.1\pm 1.4 $&$19.1\pm 1.7 $&$15.3\pm 1.4 $
\nl
4590& $13.6\pm 0.9 $&$15.5\pm 1.0 $&$17.3\pm 1.2 $&$13.2\pm 0.9 $&$16.8\pm 1.1
$&$13.1\pm 0.9 $&$16.4\pm 1.1 $&$13.0\pm 0.8 $&$16.0\pm 1.0 $&$12.9\pm 0.8 $
\nl
5024& $15.4\pm 1.4 $&$17.6\pm 1.6 $&$19.6\pm 1.8 $&$15.0\pm 1.4 $&$19.2\pm 1.7
$&$14.9\pm 1.3 $&$18.7\pm 1.7 $&$14.8\pm 1.3 $&$18.4\pm 1.7 $&$14.8\pm 1.3 $
\nl
5053& $15.5\pm 1.3 $&$17.4\pm 1.5 $&$19.7\pm 1.7 $&$15.2\pm 1.3 $&$18.9\pm 1.6
$&$14.9\pm 1.2 $&$18.2\pm 1.6 $&$14.5\pm 1.2 $&$17.4\pm 1.5 $&$14.1\pm 1.2 $
\nl
5272& $14.1\pm 0.8 $&$16.4\pm 1.0 $&$17.9\pm 1.1 $&$13.6\pm 0.8 $&$17.9\pm 1.1
$&$13.8\pm 0.8 $&$17.8\pm 1.1 $&$14.0\pm 0.8 $&$17.8\pm 1.1 $&$14.2\pm 0.8 $
\nl
5466& $16.5\pm 1.9 $&$18.7\pm 2.0 $&$21.0\pm 2.4 $&$16.1\pm 1.9 $&$20.3\pm 2.1
$&$15.9\pm 1.7 $&$19.7\pm 1.9 $&$15.6\pm 1.5 $&$19.1\pm 1.7 $&$15.4\pm 1.4 $
\nl
5897& $15.1\pm 1.8 $&$17.5\pm 2.1 $&$19.2\pm 2.3 $&$14.6\pm 1.8 $&$19.1\pm 2.3
$&$14.7\pm 1.8 $&$19.0\pm 2.3 $&$14.9\pm 1.8 $&$19.0\pm 2.3 $&$15.2\pm 1.8 $
\nl
5904& $14.6\pm 1.1 $&$17.2\pm 1.3 $&$18.6\pm 1.4 $&$14.0\pm 1.0 $&$18.8\pm 1.4
$&$14.4\pm 1.1 $&$19.0\pm 1.4 $&$14.8\pm 1.1 $&$19.2\pm 1.4 $&$15.3\pm 1.1 $
\nl
6101& $12.5\pm 1.2 $&$14.4\pm 1.3 $&$15.9\pm 1.5 $&$12.1\pm 1.1 $&$15.7\pm 1.5
$&$12.2\pm 1.1 $&$15.5\pm 1.4 $&$12.3\pm 1.1 $&$15.4\pm 1.4 $&$12.4\pm 1.1 $
\nl
6121& $15.5\pm 1.7 $&$18.3\pm 2.0 $&$19.6\pm 2.1 $&$14.8\pm 1.6 $&$19.9\pm 2.1
$&$15.3\pm 1.6 $&$20.2\pm 2.1 $&$15.8\pm 1.7 $&$20.6\pm 2.2 $&$16.4\pm 1.7 $
\nl
6171& $15.7\pm 1.9 $&$18.8\pm 2.3 $&$19.9\pm 2.4 $&$15.0\pm 1.8 $&$20.5\pm 2.5
$&$15.7\pm 1.9 $&$21.2\pm 2.5 $&$16.6\pm 2.0 $&$21.9\pm 2.6 $&$17.4\pm 2.1 $
\nl
6205& $16.1\pm 2.2 $&$18.8\pm 2.6 $&$20.5\pm 2.8 $&$15.5\pm 2.1 $&$20.5\pm 2.8
$&$15.8\pm 2.2 $&$20.4\pm 2.8 $&$16.0\pm 2.2 $&$20.4\pm 2.8 $&$16.3\pm 2.2 $
\nl
6218& $12.0\pm 1.1 $&$14.1\pm 1.3 $&$15.2\pm 1.4 $&$11.5\pm 1.0 $&$15.4\pm 1.4
$&$11.9\pm 1.1 $&$15.6\pm 1.4 $&$12.2\pm 1.1 $&$15.9\pm 1.5 $&$12.7\pm 1.1 $
\nl
6254& $17.4\pm 1.7 $&$20.4\pm 2.0 $&$22.2\pm 2.2 $&$16.8\pm 1.7 $&$22.2\pm 2.1
$&$17.1\pm 1.7 $&$22.2\pm 2.1 $&$17.4\pm 1.7 $&$22.3\pm 2.1 $&$17.8\pm 1.7 $
\nl
6341& $19.4\pm 1.5 $&$22.1\pm 1.7 $&$24.8\pm 1.9 $&$19.0\pm 1.4 $&$24.0\pm 1.8
$&$18.7\pm 1.4 $&$23.3\pm 1.7 $&$18.4\pm 1.4 $&$22.6\pm 1.7 $&$18.2\pm 1.4 $
\nl
6352& $12.9\pm 1.1 $&$16.3\pm 1.3 $&$16.8\pm 1.3 $&$12.0\pm 1.0 $&$17.9\pm 1.4
$&$13.2\pm 1.1 $&$19.3\pm 1.5 $&$14.5\pm 1.2 $&$20.6\pm 1.5 $&$15.9\pm 1.3 $
\nl
6397& $18.4\pm 1.7 $&$21.2\pm 1.9 $&$23.4\pm 2.2 $&$17.8\pm 1.7 $&$23.0\pm 2.1
$&$17.8\pm 1.6 $&$22.7\pm 2.0 $&$17.8\pm 1.6 $&$22.4\pm 2.0 $&$18.0\pm 1.6 $
\nl
6535& $16.3\pm 2.1 $&$18.9\pm 2.4 $&$20.8\pm 2.6 $&$15.8\pm 2.0 $&$20.6\pm 2.6
$&$15.9\pm 2.0 $&$20.4\pm 2.5 $&$16.0\pm 2.0 $&$20.4\pm 2.5 $&$16.3\pm 2.0 $
\nl
6584& $12.7\pm 1.2 $&$14.9\pm 1.4 $&$16.2\pm 1.6 $&$12.3\pm 1.2 $&$16.2\pm 1.5
$&$12.5\pm 1.2 $&$16.3\pm 1.5 $&$12.8\pm 1.2 $&$16.4\pm 1.5 $&$13.1\pm 1.2 $
\nl
6652& $ 9.1\pm 1.3 $&$11.4\pm 1.7 $&$12.2\pm 1.8 $&$ 8.5\pm 1.2 $&$12.8\pm 1.8
$&$ 9.1\pm 1.3 $&$13.5\pm 1.9 $&$ 9.8\pm 1.4 $&$14.1\pm 1.9 $&$10.5\pm 1.5 $
\nl
6752& $17.7\pm 1.9 $&$20.7\pm 2.2 $&$22.5\pm 2.3 $&$17.0\pm 1.8 $&$22.5\pm 2.3
$&$17.4\pm 1.8 $&$22.6\pm 2.3 $&$17.7\pm 1.8 $&$22.8\pm 2.3 $&$18.2\pm 1.9 $
\nl
6809& $16.5\pm 1.2 $&$19.1\pm 1.3 $&$21.0\pm 1.5 $&$16.0\pm 1.2 $&$20.8\pm 1.4
$&$16.1\pm 1.1 $&$20.5\pm 1.4 $&$16.1\pm 1.1 $&$20.4\pm 1.3 $&$16.3\pm 1.1 $
\nl
6838& $11.3\pm 0.9 $&$14.3\pm 1.1 $&$14.8\pm 1.1 $&$10.5\pm 0.8 $&$15.8\pm 1.1
$&$11.5\pm 0.9 $&$16.9\pm 1.2 $&$12.5\pm 0.9 $&$18.0\pm 1.3 $&$13.7\pm 1.0 $
\nl
7006& $14.0\pm 1.1 $&$16.4\pm 1.3 $&$17.9\pm 1.4 $&$13.5\pm 1.1 $&$17.9\pm 1.4
$&$13.8\pm 1.1 $&$17.9\pm 1.4 $&$14.0\pm 1.1 $&$18.0\pm 1.4 $&$14.3\pm 1.1 $
\nl
7078& $17.1\pm 1.7 $&$19.5\pm 2.0 $&$21.9\pm 2.2 $&$16.7\pm 1.7 $&$21.2\pm 2.2
$&$16.5\pm 1.7 $&$20.7\pm 2.1 $&$16.3\pm 1.7 $&$20.2\pm 2.1 $&$16.2\pm 1.7 $
\nl
7099& $16.9\pm 1.6 $&$19.3\pm 1.8 $&$21.6\pm 2.0 $&$16.5\pm 1.5 $&$21.0\pm 1.9
$&$16.3\pm 1.5 $&$20.4\pm 1.8 $&$16.1\pm 1.5 $&$19.9\pm 1.8 $&$16.0\pm 1.4 $
\nl
7492& $17.6\pm 1.9 $&$20.3\pm 2.0 $&$22.4\pm 2.4 $&$17.0\pm 1.9 $&$22.1\pm 2.2
$&$17.1\pm 1.7 $&$21.9\pm 2.1 $&$17.2\pm 1.6 $&$21.7\pm 2.0 $&$17.4\pm 1.6 $
\nl
Ter 7  & $ 7.2\pm 0.5 $&$ 8.7\pm 0.8 $&$ 9.0\pm 0.8 $&$ 6.8\pm 0.4 $&$ 9.7\pm
0.9 $&$ 7.3\pm 0.5 $&$10.6\pm 1.0 $&$ 7.9\pm 0.7 $&$11.4\pm 1.1 $&$ 8.7\pm 0.8
$ \nl
Ter 8  & $16.9\pm 1.5 $&$19.4\pm 1.7 $&$21.5\pm 1.9 $&$16.4\pm 1.5 $&$21.1\pm
1.8 $&$16.4\pm 1.4 $&$20.7\pm 1.8 $&$16.3\pm 1.4 $&$20.4\pm 1.8 $&$16.3\pm 1.4
$ \nl
Rup 106& $13.2\pm 0.8 $&$15.7\pm 0.9 $&$17.4\pm 1.0 $&$12.7\pm 0.8 $&$17.4\pm
1.0 $&$12.9\pm 0.8 $&$17.3\pm 1.0 $&$13.0\pm 0.8 $&$17.2\pm 1.0 $&$13.2\pm 0.8
$ \nl
Pal 5  & $13.4\pm 1.4 $&$15.7\pm 1.6 $&$17.0\pm 1.9 $&$12.9\pm 1.4 $&$17.1\pm
1.8 $&$13.2\pm 1.4 $&$17.2\pm 1.7 $&$13.5\pm 1.3 $&$17.4\pm 1.6 $&$13.9\pm 1.3
$ \nl
Pal 12 & $ 9.4\pm 1.4 $&$11.7\pm 1.7 $&$12.6\pm 1.9 $&$ 8.8\pm 1.4 $&$13.1\pm
1.8 $&$ 9.3\pm 1.4 $&$13.5\pm 1.8 $&$ 9.8\pm 1.4 $&$13.9\pm 1.8 $&$10.4\pm 1.4
$ \nl
IC4499 & $10.6\pm 1.1 $&$12.2\pm 1.2 $&$13.4\pm 1.4 $&$10.3\pm 1.0 $&$13.3\pm
1.3 $&$10.4\pm 1.0 $&$13.1\pm 1.3 $&$10.4\pm 1.0 $&$13.1\pm 1.3 $&$10.6\pm 1.0
$ \nl
Arp 2  & $12.3\pm 0.8 $&$14.2\pm 1.0 $&$15.6\pm 1.1 $&$11.9\pm 0.8 $&$15.5\pm
1.1 $&$12.0\pm 0.8 $&$15.4\pm 1.0 $&$12.1\pm 0.8 $&$15.4\pm 1.0 $&$12.4\pm 0.8
$ \nl
\label{tab3}
\end{planotable}
\normalsize

Among these young clusters, Ter 7 stands out with an age of $\sim 9$
Gyr, which is at least 2 Gyr younger than the others.  It appears that
Ter 7 is associated with the recently discovered Sgr dwarf spheroidal
galaxy (Ibata, Gilmore \& Irwin 1994).  Indeed Ter 7, Ter 8, Arp 2,
and M54 (not on our list) are located at {\em approximately} the same
distance and the same region of the sky as Sgr.  In addition, Da Costa
\& Armandroff (1995) have shown that the above clusters have similar
radial velocities to that of Sgr.  However, an inspection of Table 2
reveals that Ter 7 and Ter 8 have $\rgc = 14.3$ kpc while Arp 2 has
$\rgc = 21.5$ kpc.  Using the reddening, metallicity and V(HB) listed
by Da Costa \& Armandroff (1995) along with our preferred \mvrr
{}~relation, the Sgr dwarf is located at $\rgc = 15.6$ kpc.  This rather
large range in \rgc ~suggests that perhaps Ter 7, Ter 8 and Arp 2 are
not associated with Sgr. However, there are considerable uncertainties
associated with determining the Galacto-centric distances; errors in
the reddening, metallicity, the magnitude of the HB, and the
uncertainty in the correct \mvrr ~relation all lead to uncertainty in
the derived \rgc ~distances.  In this regard, we note that slightly
different choices for the input parameters, can yield Galacto-centric
distances which agree within 0.9 kpc for Ter 7, Ter 8 and Sgr, with
Arp 2 still being somewhat anomalous, with a \rgc ~value which is
about 3 kpc higher than the other objects.  While a definitive answer
will only come from proper motion studies, it appears that Ter 7, Ter
8, and Sgr are associated.  Although the evidence for Arp 2 being
associated with Sgr is not as strong, it still remains a possibility,
which should be investigated further.  Thus, two of the anomalously
young GCs were likely formed as part of Sgr, and are now being
accreted onto our Galaxy.

Lin \& Richer (1992) and Buonanno \ea (1994) have suggested that Pal
12, Arp 2, Rup 106 and Ter 7 may all have been captured by our Galaxy,
and represent later infall events.  As such, they are not indicative
of the early formation of the Galactic halo.  This argument is based
on the fact that these four clusters lie along a single great circle,
which passes through the Magellanic Stream.  A similar argument holds
for IC4499 (Fusi Pecci, Bellazzini \& Ferraro 1995).  However, even if
they have been captured by the Galaxy, their formation occured within
the halo.  It is clear that these young clusters formed much later
than the majority of GCs in the Galactic halo. Thus, it is true that
these clusters were not part of the early halo formation in the
Galaxy.  However, whether these clusters are later accretion events or
not, they are part of the Galactic halo, and so give us insights into
its formation.

\section{Age, Metallicity and Galactocentric Distance}
The question of whether or not an age-metallicity relationship exists
in the Galactic halo is a long standing problem.  It has long been
realized that if \mvrr ~has a shallow slope with metallicity, then an
age-metallicity relationship will exist, with the most metal-poor
clusters being the oldest (e.g.~Sandage 1982; Sarajedini \& King
1989).  This trend is illustrated in Figure 1, which plots the age as
a function of metallicity, assuming $\mvrr = 0.20\,\feh + 0.98$.
\begin{figure}
\vspace*{7cm}
\caption{Age as a function of metallicity for 43 GCs, assuming
$\mvrr = 0.20\,\feh + 0.98$.}
\label{figfeh}
\end{figure}
A least squares fit to all of the data, which takes into account the
error bars in both coordinates (Press \ea 1992), yields $t_9 =
(-4.0\pm 0.4)\,\feh + (9.8\pm 0.7)$, with the probability of a
non-zero slope being greater than $99.999\%$.  If the disk clusters
(47 Tuc, NGC 6352, 6838) and the GCs which are likely associated with
Sgr (Ter 7, Ter 8 and Arp 2) are removed from the fit (to leave a pure
halo sample) then the least squares solution yields $t_9 = (-4.4\pm
0.9)\,\feh +(9.3\pm 1.4)$, with the probability of a non-zero slope
being greater than $99.999\%$.  If the sample is further divided by
removing the remaining young clusters or metal-rich clusters ($\feh >
-1$; NGC 6171, 6652, Rup 106, IC4499, and Pal 12) then we find $t_9 =
(-3.4\pm 1.0)\,\feh +(11\pm 2)$, with the probability of an
age-metallicity relation existing being greater than $99.7\%$.  Note
that the above results are contingent upon a constant $[\alpha/{\rm
Fe}]$ below $\feh = -1.0$, as this was assumed in the isochrones.
However, we note that to remove the age-metallicity trend observed
would require that $[\alpha/{\rm Fe}]$ increase by 0.8 dex, from $\feh
= -2.2$ to $\feh = -1.2$, the \feh ~range spanned by the majority of
the GCs in the sample.  However, such a dramatic change in
$[\alpha/{\rm Fe}]$ is not consistent with the halo star observations,
as discussed in \S 2.1.

To further explore the age-metallicity question, GC ages have been
determined using our best estimate for the \mvrr ~zero-point, and
slopes ranging from 0.15 to 0.30, in steps of 0.002.  The resulting
ages were then analyzed using the least-squares fit as above, in order
to examine how the slope of \mvrr ~with metallicity affects the
age-metallicity relationship. The results are shown in Figure 2, which
plots the probability of an age-metallicity relation as a function of
the \mvrr ~slope with metallicity.  If all clusters are included in
the analysis, then an age-metallicity relation exists at the
greater than 99.9\% confidence level for all values of the slope
tested (from 0.15 to 0.30).  If the disk clusters and the GCs
associated with Sgr are removed from the fit, leaving a pure halo
sample, then an age-metallicity relationship exists at a greater than
97.6\% confidence level for all slopes less than or equal to 0.3.
If the young and/or metal-rich clusters are excluded, then an
age-metallicity relationship exists at the $2\,\sigma$ (95\%) level,
provided that the slope of \mvrr ~with metallicity is less than 0.259.
As most evidence favors low values for the \mvrr ~slope (see also \S
6), Figure 2 demonstrates that an age-metallicity relationship exists
in the halo of our Galaxy.
\begin{figure}
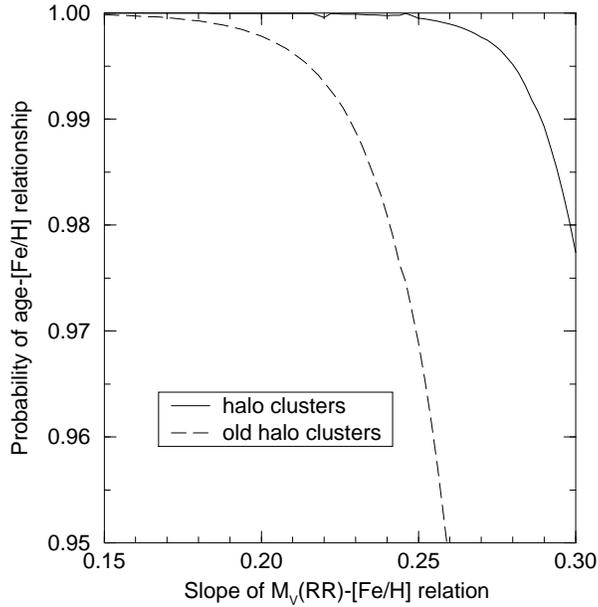

\vspace*{7cm}
\caption{The probability of an age-metallicity relation
is plotted as a function of the slope with metallicity of
\mvrr.  If all clusters are included (not plotted), then the
probability is greater than 0.999 for all values of the slope. The halo
cluster sample does not include 47 Tuc, NGC 6352, NGC 6838, Ter 7, Ter
8 and Arp 2, while the old halo sample excludes NGC 6171, 6652, Rup
106, IC4499, and Pal 12 in addition to the above clusters.  Provided
that the slope of \mvrr ~with metallicity is less than 0.26, an
age-metallicity relationship exists in the halo, regardless of which
sample is used to define the halo.}
\label{figprob}
\end{figure}

The relationship between age and Galactocentric distance is shown in
Figure 3, which plots age (assuming $\mvrr = 0.20\,\feh +
0.98$) as a function of $\rm R_{GC}$. The sample has been classified
into old halo, younger halo and disk clusters based on metallicity, kinematics
and HB morphology (Zinn 1993). This will be discussed in more detail
in section 5.  Here we note that no clear relationship between
age and Galactocentric distance exists, though there is a suggestion
that GCs become younger as one goes to large Galactocentric
distances.  A least-squares fit to the data yields $t_9 = (-0.06\pm
0.03)\,{\rm R_{GC}} + (16.5\pm 0.7)$ with non-zero slope being
significant only at the $94.5\%$ confidence level.  If the apogalactica
distances given in Table 2 are substituted for \rgc ~where available,
then the significance of the non-zero slope drops below 50\%.
Using ages and distances derived from the other HB relationships
given in Table 1 results in similar plots.  Hence, the present
data provide no compelling evidence for an age-Galactocentric distance
relationship.
\begin{figure}[t]
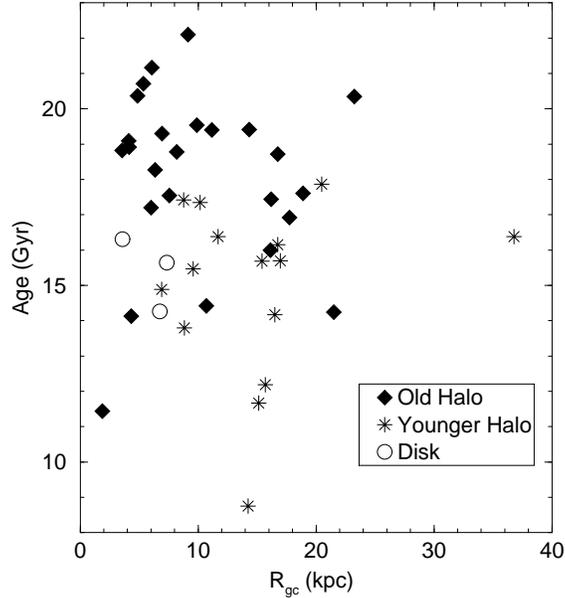

\vspace*{7cm}
\caption{Age as a function of Galactocentric distance ($\rm R_{GC}$)
for 43 GCs, assuming $\mvrr = 0.20\,\feh + 0.98$.  The sample has been
classified into old halo, younger halo and disk clusters based on metallicity,
kinematics and HB morphology (Zinn 1993).  Error bars have not
been plotted for clarity, but are typically $\pm 1.6$ Gyr. }
\label{figrgc}
\end{figure}

\section{Age Range}
In Table 3 there are a wide range of GC ages for a given \mvrr
{}~relation.  Clearly, some of this is due to the relatively large
errors (of order $\pm 1.6$ Gyr) in the individual age determinations.
To quantify how much of the age range is due to the errors, and
whether an intrinsic age range exists within the GC system, the
following statistical test was performed: an `expected' distribution
for no intrinsic age range was constructed by randomly generating
10,000 ages using a Gaussian distribution, with a mean given by the
mean age of the entire sample, and the sigma (i.e.~standard deviation)
given by the error in an individual age determination.  This is
repeated for all clusters in the sample, so that the expected
distribution contains $43\times 10,000 = 430,000$ ages.  This expected
distribution is then compared to the actual age distribution, using
the F-test (Press \ea 1992), which determines if the two distributions
have the same variance.  If there is less than a 5\% chance that the
two distributions have the same variance, then we conclude that an age
range exists.  The size of the age range is inferred by the standard
method, $\sigma_{\rm range} = \sqrt{\sigma^2_{\rm obs} - \sigma^2_{\rm
expected}}$, where $\sigma_{\rm range}$ is the sigma of the true age
range, $\sigma_{\rm obs}$ is the sigma of the actual data, and
$\sigma_{\rm expected}$ is the sigma of our expected distribution,
given the input errors in our ages.  Tests have been performed which
indicated that the typical error in our inferred $\sigma_{\rm range}$
is $\pm 0.1$ Gyr for a given \mvrr ~relation.
\begin{figure}[t]
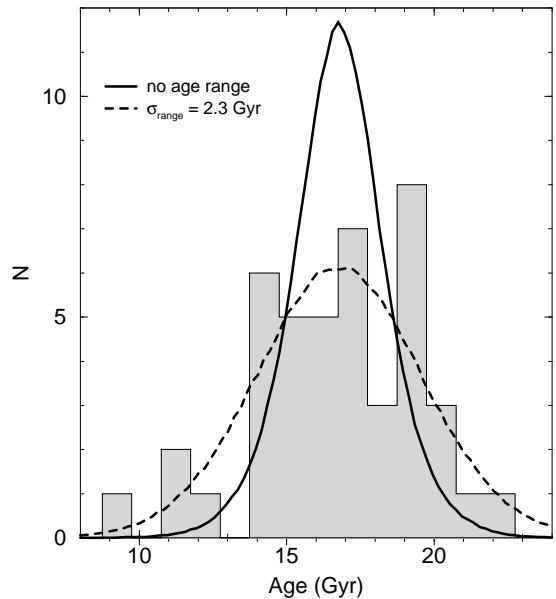

\vspace*{7cm}
\caption{Histogram of ages for our preferred \mvrr ~relation.
The solid line is the expected histogram of ages given the errors in the
individual ages, and assuming no intrinsic age range.  It has been
normalized to the total number of clusters in our sample (43).  It is clearly
not a good fit to the data, and the F-test rejects the hypothesis that
the two distributions have the same variance at a very high
confidence level.  The dotted line shows the best fitting Gaussian
distribution, which includes an intrinsic age range of $\sigma_{\rm
range} = 2.3$ Gyr, in addition to the scatter induced by the error in
the individual ages.}
\label{fighist}
\end{figure}

When this analysis is performed on all 43 clusters, the F-test rejects
the hypothesis of no intrinsic age range, at a very high confidence
level for all \mvrr ~relations used in this paper.  Figure 4 plots the
actual age histogram for our preferred \mvrr ~relation, along with the
expected histogram if there was no age range, and the best fitting
histogram, which includes an intrinsic range of ages, with
$\sigma_{\rm range} = 2.3$ Gyr.  If the age spread is defined to be
the age range which includes 95\% of the clusters, then the age spread
inferred is $4\times 2.3 = 9.2$ Gyr.  The complete results of this
analysis are presented in detail in Table 4, for all of the \mvrr
{}~relations.
\footnotesize
\begin{table}[t]
  \begin{center}
  \begin{tabular}{clrr}
\multicolumn{4}{c}{TABLE 4}\\
\multicolumn{4}{c}{AGE RANGE}\\[3pt]
\hline\hline
&\multicolumn{1}{c}{Probability of}&
\multicolumn{1}{c}{$\sigma_{\rm range}$}&
\multicolumn{1}{c}{Age}\\
\multicolumn{1}{c}{${\rm M_V(RR\,Lyr})$}&
\multicolumn{1}{c}{NO Age Range}&
\multicolumn{1}{c}{(Gyr)}&
\multicolumn{1}{c}{Spread}\\[2pt]
\hline
{}~\\[-6pt]
\multicolumn{4}{c}{All Clusters (N=43)}\\[3pt]
\hline
& & \phantom{$\sigma_{\rm range}$} & \phantom{Age Spread}  \\[-6pt]
 $0.17\feh + 0.79$   &  $2.1\times 10^{-12}$   & $2.2$   & $ 8.7$  \\
 $0.20\feh + 0.98$   &  $1.7\times 10^{-9}$    & $2.3$   & $ 9.2$  \\
 $0.15\feh + 0.98$   &  $1.8\times 10^{-11}$   & $2.7$   & $10.8$  \\
 $0.15\feh + 0.72$   &  $6.9\times 10^{-14}$   & $2.2$   & $ 8.9$  \\
 $0.20\feh + 1.06$   &  $2.6\times 10^{-9}$    & $2.4$   & $ 9.8$  \\
 $0.20\feh + 0.82$   &  $4.2\times 10^{-11}$   & $2.0$   & $ 8.1$  \\
 $0.25\feh + 1.14$   &  $2.2\times 10^{-7}$    & $2.2$   & $ 8.8$  \\
 $0.25\feh + 0.91$   &  $5.5\times 10^{-9}$    & $1.9$   & $ 7.5$  \\
 $0.30\feh + 1.22$   &  $2.3\times 10^{-6}$    & $2.1$   & $ 8.3$  \\
 $0.30\feh + 1.01$   &  $2.8\times 10^{-7}$    & $1.8$   & $ 7.0$  \\
\hline
{}~\\[-6pt]
\multicolumn{4}{c}{Excluding Young Clusters (N=39)}\\[3pt]
\hline
{}~\\[-6pt]
 $0.17\feh + 0.79$  &  $7.1\times 10^{-4}$   & $1.4$   & $ 5.7$  \\
 $0.20\feh + 0.98$  &  $9.0\times 10^{-3}$   & $1.4$   & $ 5.6$  \\
 $0.15\feh + 0.98$  &  $1.2\times 10^{-3}$   & $1.8$   & $ 7.1$  \\
 $0.15\feh + 0.72$  &  $1.2\times 10^{-4}$   & $1.5$   & $ 6.0$  \\
 $0.20\feh + 1.06$  &  $1.4\times 10^{-2}$   & $1.5$   & $ 5.9$  \\
 $0.20\feh + 0.82$  &  $3.3\times 10^{-3}$   & $1.3$   & $ 5.1$  \\
 $0.25\feh + 1.14$  &  $4.1\times 10^{-2}$   & $1.3$   & $ 5.2$  \\
 $0.25\feh + 0.91$  &  $2.0\times 10^{-2}$   & $1.1$   & $ 4.4$  \\
 $0.30\feh + 1.22$  &  $4.7\times 10^{-2}$   & $1.3$   & $ 5.1$  \\
 $0.30\feh + 1.01$  &  $4.3\times 10^{-2}$   & $1.0$   & $ 4.1$  \\
\hline
{}~\\[-6pt]
\multicolumn{4}{c}{Clusters with $-1.8 < \feh < -1.3$ (N=21)}\\[3pt]
\hline
{}~\\[-6pt]
 $0.17\feh + 0.79$  &  $1.9\times 10^{-2}$   & $1.3$   & $ 5.3$  \\
 $0.20\feh + 0.98$  &  $1.0\times 10^{-2}$   & $1.6$   & $ 6.4$  \\
 $0.15\feh + 0.98$  &  $1.9\times 10^{-2}$   & $1.7$   & $ 6.8$  \\
 $0.15\feh + 0.72$  &  $1.6\times 10^{-2}$   & $1.3$   & $ 5.1$  \\
 $0.20\feh + 1.06$  &  $1.3\times 10^{-2}$   & $1.7$   & $ 6.9$  \\
 $0.20\feh + 0.82$  &  $1.9\times 10^{-2}$   & $1.3$   & $ 5.2$  \\
 $0.25\feh + 1.14$  &  $1.0\times 10^{-2}$   & $1.8$   & $ 7.0$  \\
 $0.25\feh + 0.91$  &  $1.3\times 10^{-2}$   & $1.4$   & $ 5.4$  \\
 $0.30\feh + 1.22$  &  $7.7\times 10^{-3}$   & $1.8$   & $ 7.2$  \\
 $0.30\feh + 1.01$  &  $6.0\times 10^{-3}$   & $1.4$   & $ 5.8$  \\
\hline
\label{tab4a}
\end{tabular}
\end{center}
\end{table}
\normalsize

In looking at Figure 4, it is clear that the 4 very young clusters
(Ter 7, Pal 12, IC 4499 and NGC 6652) are somewhat anomalous, and
responsible for a good part of the very large inferred age range.  Thus,
the above analysis was repeated excluding the 4 very young clusters.
This gives a reasonable estimate of the true age range among the bulk
of the Galactic GCs.  Even with this restricted sample,
an intrinsic age range exists at the greater than the 95\% confidence
for all \mvrr ~relations (Table 4).  The size of the age range is
reduced, and varies between 4.1 --- 7.1 Gyr, depending on the choice
of \mvrr.  The sensitivity of these results to the slope of the \mvrr
{}~relation with metallicity may be reduced by considering only those
GCs in the restricted range $-1.8 < \feh < -1.3$.  There
are 21 GCs in this group (the young clusters Ter 7, Pal
12 and NGC 6652 are not included), and the results are quite similar
to those obtained with the sample which excludes the very young
clusters.  As shown in Table 4, an intrinsic age range exists
regardless of the choice of \mvrr.  The age spread is 5.1 --- 7.2,
depending on the choice of \mvrr.  Thus, we may conclude that a real
age spread of $\sim 5$ Gyr exists among the bulk of the GCs,
with several clusters ($\sim 10\%$) which are considerably younger.

\section{The Second Parameter Problem}
The morphology of the HB (ie.~the relative number of red, blue and RR
Lyr stars on the HB) is primarily governed by the metallicity of the
cluster.  As such, \feh ~is the first parameter which controls HB
morphology.  However, it has long been known that two clusters with
similar \feh ~values, can have considerably different HB morphologies.
NGC 288 and 362, and M13 and M3 are classic examples of GCs which
demonstrate that some other parameter besides \feh ~is important in
determining the morphology of the HB. Searle \& Zinn (1978)
demonstrated that the second parameter is correlated with
Galactocentric distance; there is a tight relationship between \feh
{}~and HB type in the inner halo ($\rm R_{GC} < 8\,$kpc), while the
effects of the second parameter are most pronounced in the outer halo.
The quest to determine the nature of the second parameter which
governs HB morphology has been a longstanding one in astronomy.  There
are numerous possibilities for the second parameter (age, oxygen
abundance, core rotation, mass loss on the RGB, etc.).  Given that the
previous section has demonstrated that a large intrinsic age range
exists among the GCs, we will focus here on examining the hypothesis
that age is the second parameter (Searle \& Zinn 1978; Lee, Demarque
\& Zinn 1994).

On the assumption that age is the second parameter, Zinn (1993) has
divided the halo GCs into two groups, the Old Halo (OH) and Younger
Halo (YH; these groupings are given in Table 2).  GCs were deemed to
be younger if their HB types were 0.4 redder (using the $\rm
(B-R)/(B+V+R)$ index, see footnote 3 in \S 2.3) than the typical inner
halo cluster at their metallicity.  There are 25 OH clusters in our
sample, and 15 young halo clusters.  Of the clusters which are clearly
young in our sample, IC4499, Rup 106, Pal 12 and Ter 7 are all part of
the YH grouping.  Only Arp 2 is incorrectly classified as a OH
cluster.  In addition, NGC 6652 has a a young \dv ~age, even though it
is classified as a OH cluster.  This suggests that age is the dominant
second parameter.  Indeed, the YH clusters do tend to have lower ages
than the OH clusters, as shown in Figure 3.  However, incorrect
classification of Arp 2 suggests that a third parameter affects the
HB type of some clusters.  Note that NGC 6652 ($\feh = -0.89$ and HB
type $=-1.00$) lies at the boundary of the OH and YH clusters, so
whether it belongs to the OH or YH group is uncertain.

The errors in our \dv ~ages can be rather large, thus, for the bulk of
the GCs, it is difficult to say with certainty that one particular GC
is younger than another.  This difficulty may be overcome by
determining the weighted mean age of the OH and YH groups.  The
results of this calculation are shown in Table 5.  If all of the halo
clusters are included in the sample, then the YH is 2 -- 4 Gyr younger
than the OH group, and the difference in age is significant at the
$4.8 - 7.5\,\sigma$ level.  Perhaps more importantly, if the GCs are
randomly sorted into two groups of the same size as the YH and OH
groups, age differences as large as those found between the YH and OH
groups only occur 0.5\% of the time.  Given the spread in ages found
in the previous section, this latter test ensures that the differences
in the mean ages of the two groups is not just a
coincidence.  If the Sgr clusters (Ter 7, Ter 8 and Arp 2) are
removed, then an age difference of 2.5 Gyr is found significant at the $4.3 -
4.6\,\sigma$ level.  The chance of such a large age difference
occurring in random subgroups is less than 0.5\%.  If, in addition to
the Sgr clusters, all of the young clusters are excluded from the
sample (IC4499, Pal 12, Rup 106, NGC 6652), then the difference in age
drops to 1.6 -- 2.3 Gyr (at the $3.2 - 3.5\,\sigma$ level), depending
on the choice of the \mvrr ~relation.  The chance of such a large age
difference occurring in random subgroups is less than 2.0\%.  Thus, we
see that even when the obviously young clusters are removed from the
sample, there is still a significant difference in the mean age of the
OH and YH groups.  Together, these results are consistent with the
hypothesis that age is the second parameter, and that a typical second
parameter cluster is about 2--3 Gyr younger than the clusters which
possess bluer HBs at similar metallicities.
\footnotesize
\begin{table}[t]
  \begin{center}
  \begin{tabular}{ccccc}
\multicolumn{5}{c}{TABLE 5}\\
\multicolumn{5}{c}{MEAN AGES OF THE OLD \& YOUNGER HALO}\\[3pt]
\hline\hline
&&\multicolumn{1}{c}{Younger}\\
&
\multicolumn{1}{c}{Old Halo}&
\multicolumn{1}{c}{Halo}&
\multicolumn{1}{c}{$\Delta$Age}&
\multicolumn{1}{c}{\underline {$\Delta$Age}}\\[0.2ex]
\multicolumn{1}{c}{${\rm M_V(RR\,Lyr})$}&
\multicolumn{1}{c}{Age (Gyr)}&
\multicolumn{1}{c}{Age (Gyr)}&
\multicolumn{1}{c}{(Gyr)}&
\multicolumn{1}{c}{$\sigma$}\\[2pt]
\hline
{}~\\[-6pt]
\multicolumn{5}{c}{All Halo Clusters}\\
\hline
 $0.17\feh + 0.79$& $14.9\pm 0.3$& $11.6\pm 0.3$& $ 3.3\pm 0.4$& $7.5$ \\
 $0.20\feh + 0.98$& $17.4\pm 0.4$& $14.2\pm 0.4$& $ 3.2\pm 0.5$& $6.2$ \\
 $0.15\feh + 0.98$& $19.1\pm 0.4$& $15.3\pm 0.4$& $ 3.8\pm 0.6$& $6.6$ \\
 $0.15\feh + 0.72$& $14.4\pm 0.3$& $10.8\pm 0.3$& $ 3.6\pm 0.4$& $8.3$ \\
 $0.20\feh + 1.06$& $19.0\pm 0.4$& $15.7\pm 0.4$& $ 3.3\pm 0.6$& $5.9$ \\
 $0.20\feh + 0.82$& $14.6\pm 0.3$& $11.6\pm 0.3$& $ 3.0\pm 0.4$& $7.0$ \\
 $0.25\feh + 1.14$& $18.9\pm 0.4$& $15.9\pm 0.4$& $ 3.0\pm 0.6$& $5.4$ \\
 $0.25\feh + 0.91$& $14.8\pm 0.3$& $12.2\pm 0.3$& $ 2.6\pm 0.4$& $6.0$ \\
 $0.30\feh + 1.22$& $18.8\pm 0.4$& $16.2\pm 0.4$& $ 2.6\pm 0.5$& $4.8$ \\
 $0.30\feh + 1.01$& $15.0\pm 0.3$& $12.7\pm 0.3$& $ 2.3\pm 0.4$& $5.3$ \\
\hline
{}~\\[-6pt]
\multicolumn{5}{c}{Excluding Sgr Clusters}\\[3pt]
\hline
 $0.17\feh + 0.79$& $15.2\pm 0.3$& $13.0\pm 0.3$& $ 2.2\pm 0.5$& $4.6$ \\
 $0.20\feh + 0.98$& $17.8\pm 0.4$& $15.3\pm 0.4$& $ 2.5\pm 0.5$& $4.5$ \\
 $0.15\feh + 0.98$& $19.5\pm 0.4$& $16.7\pm 0.4$& $ 2.7\pm 0.6$& $4.5$ \\
 $0.15\feh + 0.72$& $14.6\pm 0.3$& $12.5\pm 0.3$& $ 2.1\pm 0.5$& $4.6$ \\
 $0.20\feh + 1.06$& $19.4\pm 0.4$& $16.7\pm 0.4$& $ 2.7\pm 0.6$& $4.5$ \\
 $0.20\feh + 0.82$& $14.9\pm 0.3$& $12.8\pm 0.3$& $ 2.1\pm 0.5$& $4.6$ \\
 $0.25\feh + 1.14$& $19.3\pm 0.4$& $16.7\pm 0.4$& $ 2.6\pm 0.6$& $4.4$ \\
 $0.25\feh + 0.91$& $15.1\pm 0.3$& $13.0\pm 0.3$& $ 2.1\pm 0.5$& $4.6$ \\
 $0.30\feh + 1.22$& $19.2\pm 0.4$& $16.8\pm 0.4$& $ 2.5\pm 0.6$& $4.3$ \\
 $0.30\feh + 1.01$& $15.3\pm 0.3$& $13.2\pm 0.3$& $ 2.1\pm 0.5$& $4.5$ \\
\hline
{}~\\[-6pt]
\multicolumn{5}{c}{Excluding Sgr \& Very Young Clusters}\\[3pt]
\hline
 $0.17\feh + 0.79$& $15.2\pm 0.3$& $13.5\pm 0.4$& $ 1.7\pm 0.5$& $3.4$ \\
 $0.20\feh + 0.98$& $17.8\pm 0.4$& $15.7\pm 0.4$& $ 2.0\pm 0.6$& $3.4$ \\
 $0.15\feh + 0.98$& $19.5\pm 0.4$& $17.1\pm 0.5$& $ 2.3\pm 0.7$& $3.5$ \\
 $0.15\feh + 0.72$& $14.6\pm 0.3$& $13.0\pm 0.4$& $ 1.7\pm 0.5$& $3.4$ \\
 $0.20\feh + 1.06$& $19.4\pm 0.4$& $17.2\pm 0.5$& $ 2.2\pm 0.6$& $3.5$ \\
 $0.20\feh + 0.82$& $14.9\pm 0.3$& $13.2\pm 0.4$& $ 1.7\pm 0.5$& $3.4$ \\
 $0.25\feh + 1.14$& $19.3\pm 0.4$& $17.2\pm 0.5$& $ 2.1\pm 0.6$& $3.3$ \\
 $0.25\feh + 0.91$& $15.1\pm 0.3$& $13.4\pm 0.4$& $ 1.7\pm 0.5$& $3.3$ \\
 $0.30\feh + 1.22$& $19.2\pm 0.4$& $17.3\pm 0.5$& $ 2.0\pm 0.6$& $3.2$ \\
 $0.30\feh + 1.01$& $15.3\pm 0.3$& $13.7\pm 0.4$& $ 1.6\pm 0.5$& $3.2$ \\
\hline
\label{tab4}
\end{tabular}
\end{center}
\end{table}\normalsize

Although our results are consistent with age being the second
parameter, they cannot entirely rule out other phenomena.  For
example, we have assumed that the helium abundance is the same for all
clusters of a given metallicity.  If two clusters of the same
metallicity have different helium abundances, it is possible to mimic
the effects of a more youthful \dv ~age, while reddening the HB
morphology\footnote{This example was pointed out to us by the referee,
Bruce Carney.}.  Thus, our results cannot conclusively prove that age
is the second parameter, only that our ages are consistent with age
being the second parameter.  However, we note that Lee \ea (1994) have
extensively discussed arguments against parameters besides age being
responsible for the second parameter.  They found problems with every
candidate second parameter except age.  For example, variations in the
helium abundance are ruled out by constraints set by RR Lyr periods.

As mentioned above, the second parameter is correlated with
Galactocentric distance, but, as demonstrated by Figure 3,
and discussed in the previous section, age is not correlated with
Galactocentric distance in our data.  This would appear to contradict
the conclusion that age is the second parameter.  This seeming
contradiction may be resolved by a few factors: (1) the relatively
small number of GCs in our sample coupled with the relatively large
age errors (Lee \ea 1994 have 83 GCs in their HB sample); (2) some of
our clusters are far from their apogalacticon, which tends to reduce
the size of radial gradients; and (3) in the region $\rm R_{GC} = 8 -
40\,$kpc, there is considerable scatter in the HB type-\feh
{}~correlation, with some clusters in this region having a similar HB
type to clusters in the inner halo ($\rm R_{GC} = 8\,$kpc).  This is
illustrated by the fact that beyond 8 kpc, our sample contains
13 OH clusters, and 14 YH clusters.  This suggests that while we
should find a greater range of ages in the outer halo sample, the
oldest clusters in the outer halo will have a similar age to the
oldest clusters in the inner halo.  Hence, one does not expect to find
a strong age-Galactocentric distance relationship even if age is the
second parameter.  Instead, there should be a greater age range in the
outer halo, as compared to the inner halo.  To test this hypothesis,
the age range calculations discussed in the previous section were
applied to the inner halo and outer halo sample.  When all clusters
were included, the probability of an age range existing was much
higher in the outer halo sample, as opposed to the inner halo sample.
However, the age range of the inner halo sample was nearly the same
as the outer halo sample (9.6 vs. 8.5 Gyr
for our preferred \mvrr ~relation).  The determination of the age range
of the inner clusters is rather uncertain, as there are only 13
clusters in this group, of which NGC 6652 is an obvious outlier.  If
it is removed from the inner halo sample, then there is no evidence
for an age spread among the inner halo clusters.  Hopefully, the
question of an age-Galactocentric distance relation, and whether
there is a difference in the range of ages found in the inner
and outer halo  will be resolved by more, high quality data.

\section{Discussion}
In analyzing our \dv ~ages of 43 GCs (presented in Table 3), the
following conclusions were drawn: (1) if the slope of \mvrr ~with
metallicity is less than 0.26, then an age-metallicity relationship
exists, with the most metal-poor clusters being the oldest; (2) our
data set does not contain strong evidence for an age-$\rm R_{GC}$
distance relationship; (3) independent of the choice of \mvrr ~there
is strong evidence for an age spread of 5 Gyr among the bulk of the
GCs; (4) about 10\% of the GCs are substantially younger than the
majority and including them in the total sample increases the age
range to about 9 Gyr; and (5) the mean age of the red-HB, second
parameter clusters is 2 -3 Gyr younger than normal clusters, which is
consistent with age being the second parameter.  It would appear that
conclusions (2) and (5) contradict each other, since the second
parameter is correlated with Galactocentric distance (Searle \& Zinn
1978; Lee \ea 1994). This contradiction may be resolved by noting that
the errors in the age determinations are rather large ($\sim 10\%$);
there is considerable scatter in the HB-$\rm R_{GC}$ relation; and
there is weak evidence for an age-$\rm R_{GC}$ distance relationship
(at the 94.5\% confidence limit) in our data set.

In addition to the above, we note that the GCs Ter 7, Ter 8 and Arp 2
appear to be associated with the recently discovered Sgr dwarf
spheroidal galaxy.  This is based on the fact that the above objects
are all located at similar distances and in the same region of the sky
as Sgr (Ibata \ea 1994).  In addition, these GCs have radial
radial velocities similar to that of Sgr (Da Costa \& Armandroff
1995).  It appears that the Galaxy is in the process of accreting the
Sgr dwarf and its accompanying GCs, two of which (Ter 7 and Arp 2) are
anomalously young compared to the bulk of the GCs in the Galaxy.

The above conclusions strengthen the original proposal by Searle \&
Zinn (1978) that the outer halo of the Galaxy formed in a slow, rather
chaotic collapse, with the Galaxy accreting material over several Gyr.
The Sgr dwarf, and associated GCs is an example of a large gas
fragment which collapsed, and self-enriched and is now being accreted
by the Galaxy.  However, not all of the outer halo ($\rm R_{GC} >
8\,$kpc) formed via {\em later} accretion.  A significant fraction
(50\%) of the outer halo clusters in our sample do not have a strong
second parameter effect, and so are part of Zinn's (1993) Old Halo
group. These old, outer halo objects were formed during the prompt
collapse of the proto-Galactic cloud, though they still may have been
accreted at a later time.  As time passed, more metal-rich GCs formed,
and were accreted into the outer halo, leading to the observed
age-metallicity relationship.

In contrast to our earlier work on the ages of 32 GCs (Chaboyer \ea
1992), the present GC sample contains no evidence that the inner halo
formed in a rapid collapse (see Figure 3).  However, if NGC 6652 is
removed from the sample, then there is {\em no evidence} for an
intrinsic age range among the inner halo clusters, which would suggest
that the inner halo did indeed form in a rapid collapse.  In contrast,
the conclusion that a large age range exists among the outer halo GCs
is a robust statement.  A more definitive answer on whether or not the
inner halo formed in a rapid collapse requires more high quality data
of inner halo clusters.

\acknowledgments We would like to thank the referee, B.~Carney. His
suggestions have lead to a substantially improved paper.  In addition,
R.~Zinn provided us with useful comments on our initial draft.  We are
grateful to R.~Zinn for providing us with a list of globular cluster
groups, and Y.-W.~Lee for allowing us to use his HB models, both in
advance of publication.  P.D.~acknowledges partial support from NASA
grants NAG5-1486 and NAG5-2795.

\appendix
\section{Estimating the Gaussian Error in \dv}
The errors in \dv ~given in Table 2 are those quoted by the observers,
who rarely specify how they have determined the error bar.  For the
statistical analysis which comprises the bulk of the paper, Gaussian
1-$\sigma$ error bars are required.  Perhaps the best way to estimate the
Gaussian error in a measurement is to analyze repeated observations.
In this vein, we have searched the literature for independent
measurements of \dv ~for which an error is also included.  The
results of our literature search are presented in Table 6, which gives
the cluster name, \dv ~value and the reference.  Using the \dv ~values
and errors, we have calculated and tabulated the quantity
$\delta \equiv (\dvtwo_a - \dvtwo_b)/(\epsilon_a^2 + \epsilon_b^2)^{1/2}$,
where $\dvtwo_a$ is the measured \dv ~with its error ($\epsilon_a$) as
reported by observer $a$, and $\dvtwo_b$ and $\epsilon_b$ are the same
quantities reported by observer $b$.  Essentially, $\delta$ is
simply the difference in the \dv ~observations, normalized by the
quoted errors.  If the observers are quoting Gaussian 1-$\sigma$ error
bars, then $\delta$ should have a Gaussian distribution, with $\sigma
= 1$.  There are 17 measurements of $\delta$ in Table 6; for
this sample size one would expect 5 values of $\delta$ in excess of 1
for Gaussian errors.  However, this occurs only twice.  This suggests
that the reported errors are an overestimate of the Gaussian one-$\sigma$ error
bars.  Indeed, the F-test (Press \ea 1992) finds that there is only a
2\% chance that $\delta$ has a standard deviation of 1.0.  The
quantity $\delta$ has an actual standard deviation of 0.61.
Thus, multiplying the quoted errors by 0.61 yields the best estimate
the Gaussian 1-$\sigma$ error in \dv.
\footnotesize
\begin{table}[t]
  \begin{center}
  \begin{tabular}{rrrl}
\multicolumn{4}{c}{TABLE 6}\\
\multicolumn{4}{c}{INDEPENDANT \dv ~OBSERVATIONS}\\[3pt]
\hline\hline
\multicolumn{1}{c}{Cluster}&
\multicolumn{1}{c}{\dv}&
\multicolumn{1}{c}{$\delta$}&
\multicolumn{1}{c}{Reference}\\[2pt]
\hline
NGC 104& $3.61\pm 0.10$&&          Table 2 \\
       & $3.81\pm 0.18$& $-0.971$&  CSD \\
       & $3.76\pm 0.10$& $-1.060$&  SK \\[5pt]

NGC 288& $3.73\pm 0.12$&&  Table 2 \\
       & $3.70\pm 0.14$& $0.163$ & SK \\
       & $3.62\pm 0.10$& $0.704$& Bergbush 1993\\
       &               &        & Pound \ea 1987 \\[5pt]

NGC 1261&    $3.57\pm 0.12$&&  Table 2 \\
        &    $3.30\pm 0.14$& $0.379$&  Alcaino \ea 1992b \\[5pt]

NGC 1851&    $3.45\pm 0.10$&&  Table 2 \\
        &    $3.34\pm 0.10$& $0.778$&  CSD \\[5pt]

NGC 3201&    $3.45\pm 0.21$&&   Brewer \ea 1993 \\
        &    $3.44\pm 0.12$& $0.041$& Alcaino \ea 1989\\
        &                  &        & Cacciari 1984 \\[5pt]

NGC 4590&    $3.42\pm 0.10$&&  Table 2 \\
        &    $3.49\pm 0.12$& $-0.448$ &   CSD\\
        &    $3.42\pm 0.10$& $0.000$ &  Alcaino \ea 1990\\
        &                  &         & Harris 1975 \\[5pt]

NGC 5897&    $3.60\pm 0.18$&&  Table 2 \\
        &    $3.52\pm 0.14$& $0.351$&  Ferraro \ea 1992b \\[5pt]

NGC 6121&    $3.68\pm 0.16$&&  Table 2 \\
        &    $3.52\pm 0.10$& $0.848$&  SK \\
        &    $3.45\pm 0.13$& $1.116$&  Kanatas \ea 1995 \\[5pt]

NGC 6171&    $3.75\pm 0.18$&&  Table 2 \\
        &    $3.70\pm 0.11$& $0.237$ &  Ferraro \ea 1991 \\[5pt]

NGC 6752&    $3.77\pm 0.16$&&  Table 2 \\
        &    $3.72\pm 0.14$& $0.235$ &  SK \\[5pt]

NGC 6809&    $3.66\pm 0.10$&&  Table 2 \\
        &    $3.54\pm 0.14$& $0.697$&  Alcaino \ea 1992a \\[5pt]

NGC 7492&    $3.72\pm 0.14$&&  Table 2 \\
        &    $3.61\pm 0.10$& $0.778$&  SK \\[5pt]

Rup 106 &    $3.32\pm 0.07$&&  Table 2 \\
        &    $3.27\pm 0.12$& $0.360$&  CSD \\
\hline
\end{tabular}
\end{center}
\end{table}
\normalsize

Another estimate for the Gaussian 1-$\sigma$ error in \dv ~may be
obtained by computing the standard deviation (using the small sample
formulae of Keeping 1962) for each set of \dv ~values given in Table
6.  This standard deviation is then compared to the mean \dv ~error
quoted by the observers in Table 6.  Dividing the standard deviation
by the mean error and taking the average of this ratio yields 0.55.
This is the amount by which one should multiply the quoted \dv ~errors
in order to obtain a Gaussian 1-$\sigma$ error.  This value is quite
similar to the 0.61 obtained above.  To be conservative, the value of
0.61 will be used.






\end{document}